\documentclass[12pt]{article}

\usepackage{natbib}

\usepackage[utf8]{inputenc}

\usepackage[colorlinks]{hyperref}

\usepackage{amsmath,amssymb,amsthm}

\usepackage{mathrsfs}

\usepackage{graphicx}

\usepackage{color}

\usepackage{indentfirst}

\usepackage{multirow}

\usepackage{float}
\usepackage{longtable}

\usepackage{subfig}

\usepackage{enumerate} 

\usepackage[toc,page]{appendix}

\allowdisplaybreaks

\newcommand{\blind}{0}

\begin{document}
	
\def\spacingset#1{\renewcommand{\baselinestretch}%
	{#1}\small\normalsize} \spacingset{1}


\if0\blind
{
	\title{\bf On the Surprising Explanatory Power of Higher Realized Moments in Practice}
	\author{Keren Shen\thanks{
		Corresponding Author: rayshenkr@hku.hk}\hspace{.2cm}\\
		Department of Statistics and Actuarial Science\\
		The University of Hong Kong\\
		\vspace{.2cm}
		Pokfulam, Hong Kong\\		
		Jianfeng Yao \\
		Department of Statistics and Actuarial Science\\
		The University of Hong Kong\\
		\vspace{.2cm}
		Pokfulam, Hong Kong\\
		Wai Keung Li \\
		Department of Statistics and Actuarial Science\\
		The University of Hong Kong\\
		\vspace{.2cm}
		Pokfulam, Hong Kong\\}
	\maketitle
} \fi

\if1\blind
{
	\bigskip
	\bigskip
	\bigskip
	\begin{center}
		{\LARGE\bf On the Surprising Explanatory Power of Higher Realized Moments in Practice}
	\end{center}
	\medskip
} \fi

\smallskip
\begin{abstract}
	Realized moments of higher order computed from intraday returns are introduced in recent years. The literature indicates that realized skewness is an important factor in explaining future asset returns. However, the literature mainly focuses on the whole market and on the monthly or weekly scale. In this paper, we conduct an extensive empirical analysis to investigate the forecasting abilities of realized skewness and realized kurtosis towards individual stock's future return and variance in the daily scale. It is found that realized kurtosis possesses significant forecasting power for the stock's future variance. In the meanwhile, realized skewness is lack of explanatory power for the future daily return for individual stocks with a short horizon, in contrast with the existing literature.
\end{abstract}

\noindent%
\vfill

	\newpage
	\spacingset{1.45} 
	\section{Introduction}
	\noindent
	Realized moments of higher order computed from intraday returns are introduced in recent years. This article conducts an extensive empirical analysis to investigate the forecasting abilities of realized skewness and realized kurtosis towards the stock's future return and variance. It is found that realized kurtosis possesses significant forecasting power for the stock's future variance.\\
	
	\noindent
	\citet{N12} firstly introduces {\em realized skewness} of the asset price returns. \citet{ACJV13} further defines {\em realized kurtosis}. Realized moments defined in the above papers are constructed by the empirical sum of the corresponding powers of returns, which we call the naive estimator. However, the naive estimator is consistent only in the absence of microstructure noise, which must be dealt with other more sophisticated approaches.\\
	
	\noindent
	Based on the pre-averaging method in \citet{JPV10} for constructing realized variance, \citet{LWL13} introduces the pre-averaging estimator for realized skewness and kurtosis. In addition, they prove the consistency of the estimators in the presence of mircostructure noise. They also find that realized skewness of the market price has significant forecasting power for the one-month-ahead excess equity market returns, by evidence from both in-sample and out-of-sample analysis. In \citet{ACJV13}, the authors investigate whether realized skewness and realized kurtosis are informative for next week's stock returns. They find that realized skewness has a significant negative effect on future stock returns. The authors also demonstrate the significance in economic sense that buying stocks with lowest realized skewness and selling stocks with highest realized skewness generates a profit significantly. In addition, realized kurtosis exhibits a positive relationship with the weekly returns, even though the evidence is not always robust and statistically significant. This motivates the question of whether the same conclusion could be applied to the individual stock and for a shorter period, say one day, as people often focus more on the short-term profit in the stock market nowadays. Continuing exploration along the line, we investigate in this paper whether higher realized moments have explaining power on variances of future assets, as estimated by realized variance.\\
	
	\noindent
	In empirical study, we show that in contrast with \citet{ACJV13} and \citet{LWL13}, realized skewness does not show enough explanatory power for future daily returns. We conduct regression analysis towards $50$ randomly selected stocks from different industries and capitalization sizes. We find that the forecasting ability of realized skewness is statistically significant for only $8$ out of $50$ stocks. On the other hand, realized kurtosis, which is able to reflect the price jump size, shows strong evidence of forecasting power for future realized variances. Thirty two out of the same $50$ stocks are shown to have the property. Moreover, we find that the square root of realized kurtosis has an even better forecasting ability for future realized variances.\\
	
	\noindent
	In addition, we compare the forecasting ability of realized kurtosis with other well-known variables, which may help in predicting asset's volatility, namely trading volume and signed daily return. In \citet{CF06}, the authors conduct regression analysis of realized volatility against trading volume, trading frequency, average trading size and order imbalance. Trading volume is comprised of trading frequency and average trading size, while order imbalance is the difference between the number of trades initiated by buyers and sellers. The authors find that daily trading volume and trading frequency give equally good predictions on realized volatility, while average trading size and order imbalance adds little explaining power. Therefore, in our empirical analysis, we only include daily trading volume as a possible covariate. Furthermore, signed returns are also informative for the volatility, especially for negative returns, which is usually interpreted as the {\em leverage effect}. The effect is firstly discussed by \citet{B76} and \citet{C82}, and is due to the reason that a negative return leads to an increase in the debt-to-equity ratio, resulting in an increase in the future volatility of the return \citep{BZ06}. In this article, we include both positive and negative daily returns as covariates. We find that in the presence of trading volume and signed daily returns, realized kurtosis still exhibits nice predicting power, in general.\\
	
	\noindent
	In summary, the main findings of the paper are the following:
	\begin{itemize}
		\item Realized skewness is lack of explanatory power for the future daily return for individual stocks with a short horizon.
		\item Realized kurtosis exhibits significant forecasting power for the future realized variance.
		\item Realized kurtosis incorporates some information contained in trading volume.
		\item There may be some nonlinear relationships between realized kurtosis and the future daily volatility.
	\end{itemize}
	
	\noindent
	The rest of the paper is organized as follows. Section $2$ reviews the estimators of higher realized moments. In Section $3$, we examine the forecasting ability of higher realized moments for the future daily return and return variance, for a chosen stock. The robustness of the result in Section 3 is checked in Section $4$. Section $5$ concludes.\\
	
	\section{Methodology}
	\subsection{Model setup}
	\noindent
	Define an adapted process $\{X_t,t \ge 0\}$ on some probability space $(\Omega,\mathcal{F},P)$ as follows: 
	\begin{equation}
	X_t = \int_{0}^{t}\mu_sds + \int_{0}^{t}\sigma_sdW_s + \sum_{s \le t} \Delta_sX, 
	\end{equation}
	where $\{\mu_s, 0 \le s \le t\}$ is an adapted locally bounded process, $\{ \sigma_s,0 \le s \le t\}$ is a c$\grave{a}$dl$\grave{a}$g volatility process, and $\Delta_sX = X_s -X_{s-}$ is the jump of $X$ at time $s$. Assume that the jump of $X$ arrives through a finite jump process, for example, the compound Poisson process. The {\em quadratic variation} for the $T$-th day is defined as:
	\begin{displaymath}
	[X,X]_T = \int_{T-1}^{T} \sigma_s^2 ds + \sum_{T-1 \le s \le T} (\Delta_sX)^2.
	\end{displaymath}
	
	\subsection{Naive estimator}
	\noindent
	Let the grid of observation times of the $T$-th day be given by $\mathcal{G} = \{t_0,t_1,\cdots,t_n\}$, which satisfies that
	\begin{displaymath}
	T-1 = t_0 < t_1 < \cdots < t_n = T.
	\end{displaymath}
	For simplicity, we assume that the observation point is equidistant which is frequently used in the literature., i.e. $t_i - t_{i-1} \equiv \delta$ for any $1 \le i \le n$. {\em Realized variance} ($rvar$) for the $T$-th day is defined as:
	\begin{equation}
	rvar := \sum_{i=1}^{n} (X_{t_i} - X_{t_{i-1}})^2.
	\end{equation}
	In the absence of microstructure noise, when $n$ goes to infinity, 
	\begin{equation}
	rvar \rightarrow_p [X,X]_T.
	\end{equation}
	
	\noindent
	Similarly we define {\em realized skewness} ($rskew$) and {\em realized kurtosis} ($rkurt$) for the $T$-th day as:
	\begin{equation}
	rskew := \sum_{i=1}^{n} (X_{t_i} - X_{t_{i-1}})^3,
	\end{equation}
	and
	\begin{equation}
	rkurt := \sum_{i=1}^{n} (X_{t_i} - X_{t_{i-1}})^4.
	\end{equation}
	In addition, realized skewness and realized kurtosis can be normalized as:
	\begin{equation}
	nrskew := \frac{rskew}{rvar^{3/2}},
	\end{equation}
	\begin{equation}
	nrkurt := \frac{rkurt}{rvar^2}.
	\end{equation}
	
	\noindent
	When $n$ goes to infinity, normalized realized skewness and kurtosis have the following limits in probability:
	\begin{equation}
	nrskew \rightarrow_p \frac{\sum_{T-1 \le s \le T} (\Delta_sX)^3}{(\int_{T-1}^{T} \sigma_s^2 ds + \sum_{T-1 \le s \le T} (\Delta_sX)^2)^{3/2}},
	\end{equation}
	\begin{equation}
	nrkurt \rightarrow_p \frac{\sum_{T-1 \le s \le T} (\Delta_sX)^4}{(\int_{T-1}^{T} \sigma_s^2 ds + \sum_{T-1 \le s \le T} (\Delta_sX)^2)^2}.
	\end{equation}
	We call the above realized moments as the naive ones, for instance, the naive realized skewness.\\ 
	
	\subsection{Pre-averaging estimator}
	\noindent
	In practice, it is commonly admitted that microstructure noise is inherent in the high-frequency price process so that we are not able to observe directly $X_{t_i}$, but $Y_{t_i}$, a noisy version of $X_{t_i}$ at times $i = 0, \cdots, n$. In this paper, we assume that
	\begin{equation}
	Y_{t_i} = X_{t_i} + \epsilon_{t_i},
	\end{equation}
	where $\epsilon_{t_i}$ are i.i.d. microstructure noise with mean zero and variance $\eta^2$, and $\epsilon_{t_i}$ and $X_{t_i}$ are independent with each other.\\
	
	\noindent
	To reduce the effect of microstructure noise, the pre-averaging method is used \citep{LWL13} within blocks of length $k_n$. In the $i$-th block, the pre-averaging return is constructed as 
	\begin{equation}
	\Delta_{i,k_n}^n Y(g) = \sum_{j=1}^{k_n} g(\frac{j}{k_n}) (Y_{t_{i+j}} - Y_{t_{i+j-1}}),
	\end{equation} 
	and 
	\begin{equation}
	\Delta_{i,k_n}^n \bar{Y}(g) = \sum_{j=1}^{k_n} (g(\frac{j}{k_n})-g(\frac{j-1}{k_n}))^2 (Y_{t_{i+j}} - Y_{t_{i+j-1}})^2,
	\end{equation} 
	with a non-negative piece-wise differentiable Lipschitz function $g$, satisfying $g(x) = 0$ when $x \notin (0,1)$ and $\bar{g}(p) = \int_{0}^{1} g^p(x)dx >0$. In empirical analysis, we use $g(x) = \min\{x,1-x\}$ for $0 \le x \le 1$, which is often used in the literature. As a result, the pre-averaging realized measures are constructed as follows:
	\begin{equation}
	rvar := \frac{1}{\bar{g}(2)}(\frac{1}{k_n} \sum_{i=1}^{n-k_n} (\Delta_{i,k_n}^n Y(g))^2 - \frac{1}{2k_n} \sum_{i=1}^{n-k_n} (\Delta_{i,k_n}^n \bar{Y}(g))),
	\end{equation}
	\begin{equation}
	rskew := \frac{1}{\bar{g}(3)}(\frac{1}{k_n} \sum_{i=1}^{n-k_n} (\Delta_{i,k_n}^n Y(g))^3),
	\end{equation}
	\begin{equation}
	rkurt := \frac{1}{\bar{g}(4)}(\frac{1}{k_n} \sum_{i=1}^{n-k_n} (\Delta_{i,k_n}^n Y(g))^4),
	\end{equation}
	and 
	\begin{equation}
	nrskew := \frac{rskew}{rvar^{3/2}}, \quad nrkurt := \frac{rkurt}{rvar^{2}}.
	\end{equation}
	
	\noindent
	In the presence of microstructure noise, the above pre-averaging estimators have the following limits in probability when $k_n, n \rightarrow \infty$ and $k_n/n \rightarrow 0$:
	\begin{equation}
	nrskew \rightarrow_p \frac{\sum_{T-1 \le s \le T} (\Delta_sX)^3}{(\int_{T-1}^{T} \sigma_s^2 ds + \sum_{T-1 \le s \le T} (\Delta_sX)^2)^{3/2}},
	\end{equation}
	\begin{equation}
	nrkurt \rightarrow_p \frac{\sum_{T-1 \le s \le T} (\Delta_sX)^4}{(\int_{T-1}^{T} \sigma_s^2 ds + \sum_{T-1 \le s \le T} (\Delta_sX)^2)^2}.
	\end{equation}

	\section{Empirical data analysis}
	\noindent
	The literature indicates that realized skewness is an important factor in explaining future asset returns. However, the literature mainly focuses on the whole market and on the monthly or weekly scale \citep{LWL13,ACJV13}. In this section, we test the cross-sectional forecasting performance of realized skewness for the individual stock and in the daily scale. Furthermore, we examine whether higher moments, namely realized skewness and realized kurtosis, have any explaining power for variances of the stock price.\\
	
	\subsection{Data and exploratory analysis}
	\noindent
	Our empirical analysis is based on the transaction prices from New York Stock Exchange (NYSE) Trade and Quote (TAQ) database for International Business Machines (IBM). The sample period starts from January 3, 2012 and ends on December 31, 2012 and the daily transaction records start from 10:00 to 15:30, to remove the open and close effect within one day. We have totally 250 days for the stock. We only report the result for the pre-averaging estimator of the stock to save space, as the conclusion is generally the same in the naive estimator case, which is computed by the summation of the corresponding power of 5-minute log-returns. We check the robustness of the result of this section by exploring other stocks in Section $4$. We find that the properties we see in this section apply in general.\\
	
	\noindent
	We firstly conduct data cleaning with the procedures introduced in \citet{BG06} and \citet{BHLS09}. The steps are the following:\\
	\begin{enumerate}
		\item Delete entries with a time stamp outside 9:30 am - 4:00 pm when the exchange is open.
		\item Delete entries with a time stamp inside 9:30 - 10:00 am or 3:30 - 4:00 pm to eliminate the open and end effect of price fluctuation.
		\item Delete entries with the transaction price equal to zero.
		\item If multiple transactions have the same time stamp, use the median price.
		\item Delete entries with prices which are outliers. Let $\{p_i\}_{i=1}^N$ be an ordered tick-by-tick price series. We call the $i$-th price an outlier if $|p_i-\bar{p}_i(m)|>3s_i(m)$, where $\bar{p}_i(m)$ and $s_i(m)$ denote the sample mean and sample standard deviation of a neighborhood of $m$ observations around $i$, respectively. For the beginning prices which may not have enough left hand side neighbors, we get $m-i$ neighbors from $i+1$ to $m+1$. Similar procedures are taken for the ending prices. We take $m=5$ here.  
	\end{enumerate}
	
	\noindent
	Daily returns are computed as the difference of the logrithm of the closing prices for the current and previous days. Realized moments are estimated by the pre-averaging method. We take $\Delta_n=$ 1 minute, $k_n = 10$ and $g(x) = \min(x,1-x)$.\\
	
\begin{figure}[H]
		\centering
		\includegraphics[width=11cm]{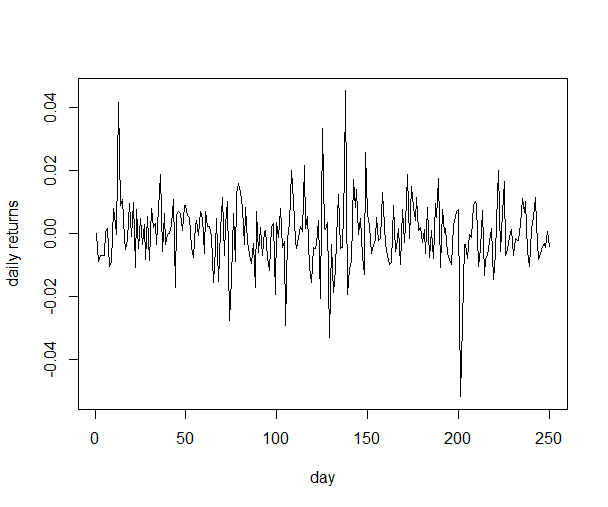}
		\caption{Daily log-returns of International Business Machines.}
		\label{f1}
	\end{figure}

\begin{figure}[H]
		\centering
		\includegraphics[width=11cm]{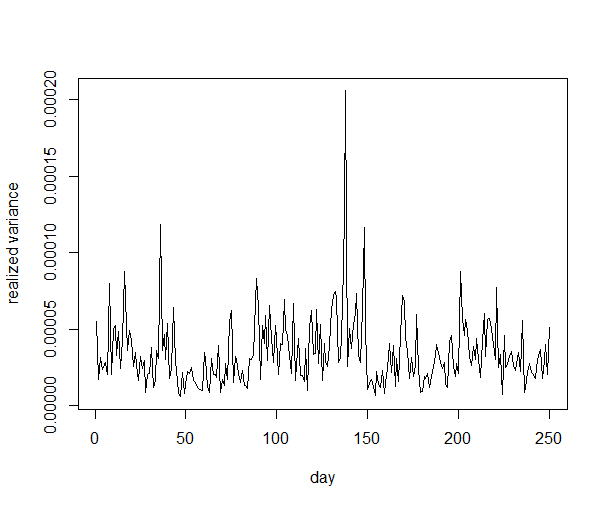}
		\caption{Daily realized variance of International Business Machines.}
		\label{f2}
	\end{figure}

\begin{figure}[H]
		\centering
		\includegraphics[width=11cm]{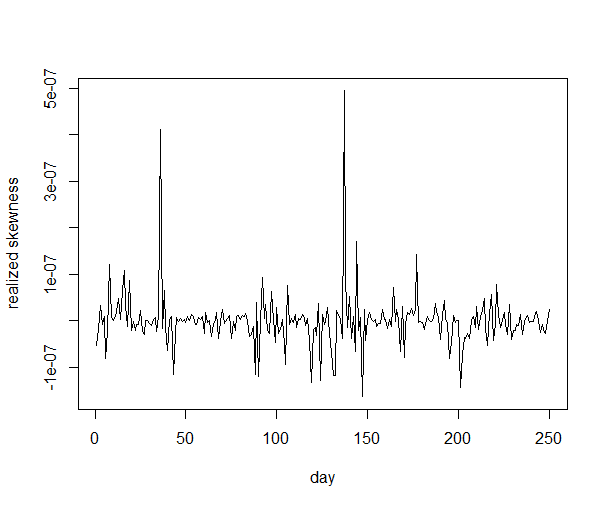}
		\caption{Daily realized skewness of International Business Machines.}
		\label{f3}
	\end{figure}

\begin{figure}[H]
		\centering
		\includegraphics[width=11cm]{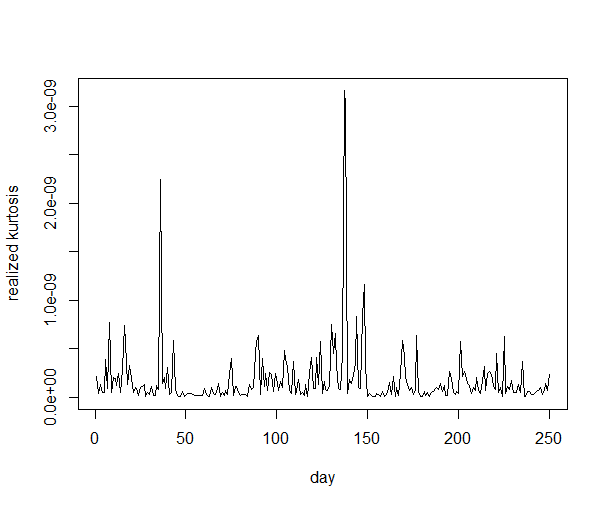}
		\caption{Daily realized kurtosis of International Business Machines.}
		\label{f4}
	\end{figure}

\begin{figure}[H]
		\centering
		\includegraphics[width=10.5cm]{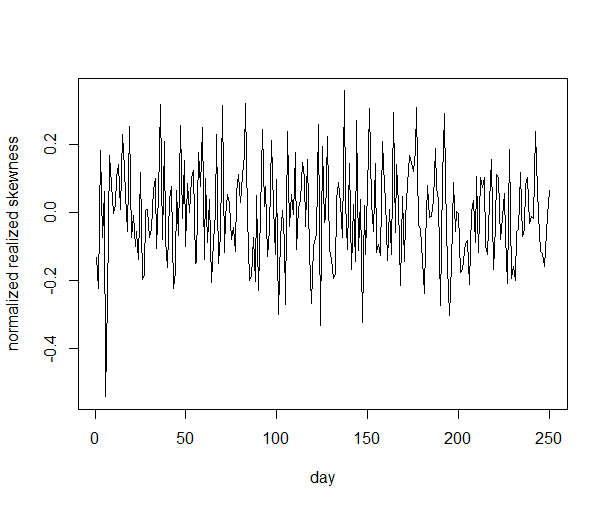}
		\caption{Daily normalized realized skewness of International Business Machines.}
		\label{f5}
	\end{figure}

\begin{figure}[H]
		\centering
		\includegraphics[width=10.5cm]{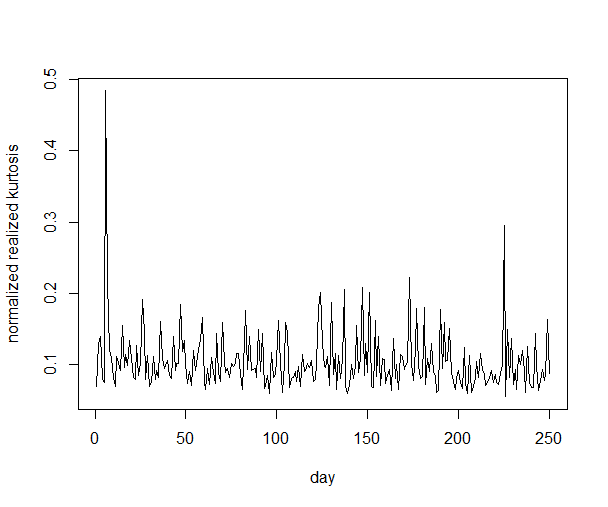}
		\caption{Daily normalized realized kurtosis of International Business Machines.}
		\label{f6}
	\end{figure}

	\noindent
	The descriptive statistics for IBM daily returns ($dret$), realized variance ($rvar$), realized skewness ($rskew$), realized kurtosis ($rkurt$), normalized realized skewness ($nrskew$) and normalized realized kurtosis ($nrkurt$) is shown in Table $\ref{t1}$. In addition, their plots are shown in Figures $\ref{f1}-\ref{f6}$.\\
	\begin{table}[H]
		\centering
		\caption{Descriptive statistics for the daily return and realized moments}
		
		\begin{tabular}{l*{6}{c}}
			\hline
			& $dret$ & $rvar$ & $rskew$ & $rkurt$ & $nrskew$  & $nrkurt$  \\
			\hline
			Maximum & 4.52$\times 10^{-2}$ & 2.06$\times 10^{-4}$ & 4.94$\times 10^{-7}$ & 3.16$\times 10^{-9}$ & 0.36 & 0.48   \\
			Minimum & -5.17$\times 10^{-2}$ & 5.82$\times 10^{-6}$ & -1.64$\times 10^{-7}$ & 5.21$\times 10^{-12}$ & -0.54 & 0.06   \\
			Mean  & 6.36$\times 10^{-5}$ & 3.47$\times 10^{-5}$ & -6.68$\times 10^{-10}$ & 1.83$\times 10^{-10}$ & -0.005 & 0.10   \\
			SD  & 1.06$\times 10^{-2}$ & 2.29$\times 10^{-5}$ & 5.80$\times 10^{-8}$ & 3.45$\times 10^{-10}$ & 0.14 & 0.04   \\
			Skewness & -3.33$\times 10^{-2}$ & 2.54 & 3.72 & 5.89 & 0.02 & 3.67   \\
			Kurtosis & 7.14 & 15.8 & 33.1 & 46.0 & 3.10 & 27.8   \\
			\hline		
		\end{tabular}
		\label{t1}
	\end{table}
	
	
	\noindent
	In Table $\ref{t1}$, we find that the daily return exhibits no significant skewness, as the p-value of the D'Agostino skewness test is $0.88$, while all other measures show positive skewness. All measures show larger kurtosis than the normal distribution, i.e. fat tails. In addition, from Figures $\ref{f1}$ to $\ref{f6}$, we can see that realized variance exhibits clear volatility clustering phenomenon and it seems that realized kurtosis shows similar pattern. In the meanwhile, realized skewness, normalized realized skewness and normalized kurtosis seem more random and behave like white noise. We conduct some basic time series analysis for these series.\\
	
	\noindent
	We fit simple time series to the pre-averaging realized variance, realized skewness, realized kurtosis, and etc. From the auto-correlation function plots and also by the Ljung-Box test (Table $\ref{ljung}$), we conclude that realized skewness, normalized realized skewness and normalized realized kurtosis can be treated as white noise.\\
	\begin{table}[H]
		\centering
		\caption{Ljung-Box test of the series with lag $10$}
		\begin{tabular}{l*{5}{c}}
			\hline
			& $rvar$ & $rskew$ & $rkurt$ & $nrskew$ & $nrkurt$\\
			\hline
			p-value & 1.78$\times 10^{-14}$ & 0.70 & 8.52$\times 10^{-9}$ & 0.41 & 0.85  \\
			\hline		
		\end{tabular}
		\label{ljung}
	\end{table}
	
	\noindent
	Furthermore, realized kurtosis can be fitted by an AR(1) with GARCH(1,1) model (with AR coefficient $0.38$ and GARCH coefficient $(0.05,0.05)$). Realized variance can be fitted by an AR(1) with ARCH(1) model (with AR coefficient $0.36$, and ARCH coefficient $0.31$).  The fitting result validates that only realized variance and realized kurtosis possess some memory.\\
	
	\subsection{Predicting daily returns}
	\noindent
	As mentioned earlier, realized skewness has been thought to have explaining power for the future daily return of the equity market. Now, we employ the regression models to investigate whether the conclusion holds for the individual stock and the daily horizon. Here, we regress daily returns with respect to the previous day realized variance, realized skewness and realized kurtosis (and their normalized counterparts).\\
	
	\noindent
	We employ the following regression models:
	\begin{equation}
	dret_{t+1} = \alpha_0 + \alpha_1 rvar_t + \alpha_2 rskew_t + \alpha_3 rkurt_t + \epsilon_{t+1}
	\end{equation}
	\begin{equation}
	dret_{t+1} = \alpha_0 + \alpha_1 rvar_t + \alpha_2 nrskew_t + \alpha_3 nrkurt_t + \epsilon_{t+1}
	\end{equation}
	The above equations are predictive regressive models for forecasting one-day ahead daily returns with different realized measures. Tables $\ref{t2}$ and $\ref{t3}$ show the result of the regression models. \\
	\begin{table}[H]
		\centering
		\caption{Regression model (19) for daily return with previous day realized moments}
		\begin{tabular}{l*{5}{c}}
			\hline
			& $\alpha_1(rvar)$ & $\alpha_2(rskew)$ & $\alpha_3(rkurt)$ & $R^2$ & F test\\
			\hline
			Estimate  & -6.05$\times 10^{1}$ & 1.65$\times 10^{4}$ & 4.35$\times 10^{6}$ & 0.02 &    \\
			Standard Error  & 7.20$\times 10^{1}$ & 1.61$\times 10^{4}$ & 5.18$\times 10^{6}$ &  &   \\
			p-value & 0.40 & 0.31 & 0.40 & & 0.22  \\
			\hline		
		\end{tabular}	
		\label{t2}
	\end{table}

	\begin{table}[H]
		\centering
		\caption{Regression model (20) for daily return with previous day realized moments}
		\begin{tabular}{l*{5}{c}}
			\hline
			& $\alpha_1(rvar)$ & $\alpha_2(nrskew)$ & $\alpha_3(nrkurt)$ & $R^2$ & F test\\
			\hline
			Estimate  & 9.32 & 1.97$\times 10^{-3}$ & 4.58$\times 10^{-3}$ & 0.001 &    \\
			Standard Error  & 3.24$\times 10^{1}$ & 5.22$\times 10^{-3}$ & 1.81$\times 10^{-2}$ &  &   \\
			p-value & 0.77 & 0.71 & 0.80 & & 0.96  \\
			\hline		
		\end{tabular}
		
		\label{t3}
	\end{table}
	
	\noindent
	From Tables $\ref{t2}$ and $\ref{t3}$, one may find that there is no linear relationship with daily return and realized variance, realized skewness, realized kurtosis and their normalized counterparts. This finding is different from some of the literature that realized skewness has explanatory power for daily returns. It is possible that the daily scale and the individual aspect make the disappearance of the explaining power. In addition, we fit the same regression models to those $50$ stocks mentioned in Section $4$. We find that for both naive and pre-averaging estimators, realized skewness has significant forecasting power for daily returns in only $8$ out of $50$ cases. For those stocks with large capitalization size, realized skewness shows significant effect with about $20\%$ chance; while for medium and small sizes, the chance becomes only one out of ten. From above analysis, we conclude that for individual stocks, realized skewness does not have enough forecasting power for the one-day ahead daily returns.\\

	\subsection{Predicting the variance}
	\subsubsection{Regression analysis}
	\noindent
	In this subsection, we regress realized variance against the previous day return, realized skewness, realized kurtosis and so on. We wish to see whether there exist some variables having predicting power for realized variances. We employ the following regression models.
	\begin{equation}
	rvar_{t+d} = \alpha_0 + \alpha_1 dret_t + \alpha_2 rskew_t + \alpha_3 rkurt_t + \epsilon_{t+d}
	\end{equation}
	\begin{equation}
	rvar_{t+d} = \alpha_0 + \alpha_1 dret_t + \alpha_2 nrskew_t + \alpha_3 nrkurt_t + \epsilon_{t+d}
	\end{equation}
	where $d=1$ for this subsection. The above equations are predictive regression models for forecasting one-day ahead realized variance with different realized measures. Tables $\ref{t4}$ and $\ref{t5}$ show the result of the regression models.\\
	\begin{table}[H]
		\centering
		\caption{Regression model (21) for realized variance with previous day realized moments, $d=1$}
		\begin{tabular}{l*{5}{c}}
			\hline
			& $\alpha_1(dret)$ & $\alpha_2(rskew)$ & $\alpha_3(rkurt)$ & $R^2$ & F test\\
			\hline
			Estimate  & -4.04$\times 10^{-4}$ & 7.32 & 2.85$\times 10^{4}$ & 0.199 &    \\
			Standard Error  & 1.65$\times 10^{-4}$ & 3.13$\times 10^{1}$ & 4.80$\times 10^{3}$ &  &   \\
			p-value & 0.02 $*$ & 0.82 & 1.3$\times 10^{-8}$ $***$ & & 1.85$\times 10^{-9}$  \\
			\hline		
		\end{tabular}
		\label{t4}
	\end{table}
	
	\begin{table}[H]
		\centering
		\caption{Regression model (22) for realized variance with previous day realized moments, $d=1$}
		\begin{tabular}{l*{5}{c}}
			\hline
			& $\alpha_1(dret)$ & $\alpha_2(nrskew)$ & $\alpha_3(nrkurt)$ & $R^2$ & F test\\
			\hline
			Estimate  & -1.87$\times 10^{-4}$ & 3.80$\times 10^{-7}$ & 4.11$\times 10^{-5}$ & 0.011 &    \\
			Standard Error  & 1.89$\times 10^{-4}$ & 1.30$\times 10^{-5}$ & 4.03$\times 10^{-5}$ &  &   \\
			p-value & 0.32 & 0.98 & 0.31 & & 0.55  \\
			\hline		
		\end{tabular}
		
		\label{t5}
	\end{table}
	
	\noindent
	From Tables $\ref{t4}$ and $\ref{t5}$, it is found that realized kurtosis is extremely significant in explaining future realized variances, while the daily return has significant effect as well, indicating a possible leverage effect. We will explore the effect more in later subsections. The coefficients estimated are positive for realized kurtosis, suggesting that larger price jump sizes lead to larger price fluctuations in the near future. We see that other realized measures show no forecasting power, for example, normalized realized kurtosis. This may be because that the effect of the jump size is normalized "out". \\
	
	\subsubsection{Regression analysis with longer horizon}
	\noindent
	We have seen that realized kurtosis has forecasting power for the one-day ahead realized variance and we explore whether the same conclusion holds for a longer forecasting horizon. As a result, the same regression model (21) with longer horizon are shown here, with horizons of 2-day, 5-day and 22-day, i.e. $d=2,5,22$, corresponding to a trading period of two days, one week and one month, respectively. The results are shown in Tables $\ref{t6}-\ref{t8}$.\\
	
	\begin{table}[H]
		\centering
		\caption{Regression model (21) for realized variance with previous day realized moments, $d=2$}	
		\begin{tabular}{l*{5}{c}}
			\hline
			& $\alpha_1(dret)$ & $\alpha_2(rskew)$ & $\alpha_3(rkurt)$ & $R^2$ & F test\\
			\hline
			Estimate  & -5.63$\times 10^{-5}$ & -6.34$\times 10^{1}$ & 1.31$\times 10^{4}$ & 0.035 &    \\
			Standard Error  & 1.82$\times 10^{-4}$ & 3.43$\times 10^{1}$ & 5.28$\times 10^{3}$ &  &   \\
			p-value & 0.76 & 0.07 & 0.01 $*$ & & 0.07  \\
			\hline		
		\end{tabular}
		\label{t6}
	\end{table}
	\begin{table}[H]
		\centering
		\caption{Regression model (21) for realized variance with previous day realized moments, $d=5$}	
		\begin{tabular}{l*{5}{c}}
			\hline
			& $\alpha_1(dret)$ & $\alpha_2(rskew)$ & $\alpha_3(rkurt)$ & $R^2$ & F test\\
			\hline
			Estimate  & 8.28$\times 10^{-5}$ & -3.30$\times 10^{1}$ & 9.84$\times 10^{3}$ & 0.020 &    \\
			Standard Error  & 1.84$\times 10^{-4}$ & 3.47$\times 10^{1}$ & 5.34$\times 10^{3}$ &  &   \\
			p-value & 0.65 & 0.34 & 0.07$\cdot$ & & 0.31  \\
			\hline		
		\end{tabular}
		
		\label{t7}
	\end{table}
	\begin{table}[H]
		\centering
		\caption{Regression model (21) for realized variance with previous day realized moments, $d=22$}	
		\begin{tabular}{l*{5}{c}}
			\hline
			& $\alpha_1(dret)$ & $\alpha_2(rskew)$ & $\alpha_3(rkurt)$ & $R^2$ & F test\\
			\hline
			Estimate  & 1.87$\times 10^{-4}$ & -2.09$\times 10^{1}$ & -5.69$\times 10^{3}$ & 0.016 &    \\
			Standard Error  & 1.81$\times 10^{-4}$ & 3.43$\times 10^{1}$ & 5.27$\times 10^{3}$ &  &   \\
			p-value & 0.30 & 0.54 & 0.28 & & 0.38  \\
			\hline		
		\end{tabular}
		
		\label{t8}
	\end{table}
	
	\noindent
	When the horizon for prediction becomes longer and longer, the predicting power of realized kurtosis on realized variance becomes less and less, which is really natural. We find that when $d=2$, realized kurtosis has a significant effect on the regressand; when $d=5$, the coefficient for realized kurtosis becomes marginally significant; and when $d=22$, the effect now becomes totally insignificant. Moreover, when the coefficient of realized kurtosis is at least marginally significant, the coefficient is positive, the same as in the case when $d=1$. In addition, all other measures always exhibit no explaining power for the future realized variance, which is the same result as for the short horizon. \\

	\subsubsection{Adding other covariates}
	\noindent
	In the literature, there exist some other covariates used to explain or/and forecast the price volatility, for example, trading volume of the stock within a period, and negative daily returns. Trading volume is a covariate used to explain the volatility in the finance field which tends to be larger in the case of higher volatility. In addition, negative daily returns reflect the so-called leverage effect. In this subsection, we regress realized variance against the previous day realized kurtosis, positive and negative daily returns and trading volume. We find that realized kurtosis still exhibits significant explaining power in the presence of other covariates.\\
	
	\noindent
	Let $tvol$ denote trading volume, $dret^{+}$ positive daily return, and $dret^{-}$ negative daily return. We employ the following regression models for analysis:\\
	\begin{equation}
	rvar_{t+1} = \alpha_0 + \alpha_1 tvol_t + \epsilon_{t+1}
	\end{equation}
	\begin{equation}
	rvar_{t+1} = \alpha_0 + \alpha_1 tvol_t + \alpha_2 rkurt_t + \epsilon_{t+1}
	\end{equation}
	
	\noindent
	The estimating results are shown in Tables $\ref{t10}$ and $\ref{t11}$.\\
	\begin{table}[H]
		\centering
		\caption{Regression model (23) for realized variance with previous day trading volume}	
		\begin{tabular}{l*{3}{c}}
			\hline
			& $\alpha_1(tvol)$ & $R^2$ & F test\\
			\hline
			Estimate  & 2.34$\times 10^{-12}$  & 0.026 &    \\
			Standard Error  & 1.02$\times 10^{-12}$  &  &   \\
			p-value & 0.02$*$  &  &  0.02  \\
			\hline		
		\end{tabular}
		
		\label{t10}
	\end{table}
	\begin{table}[H]
		\centering
		\caption{Regression model (24) for realized variance with previous day trading volume and realized kurtosis}
		\begin{tabular}{l*{4}{c}}
			\hline
			& $\alpha_1(tvol)$ & $\alpha_2(rkurt)$  & $R^2$ & F test\\
			\hline
			Estimate  & -1.16$\times 10^{-14}$ & 2.69$\times 10^{4}$  & 0.173 &    \\
			Standard Error  & 1.02$\times 10^{-12}$ & 4.56$\times 10^{3}$  &  &   \\
			p-value & 0.99 & 1.50$\times 10^{-8}***$  & & 7.79$\times 10^{-9}$  \\
			\hline		
		\end{tabular}	
		\label{t11}
	\end{table}
	
	\noindent
	We observe from Table $\ref{t10}$ that the previous day trading volume has a significant positive relationship with realized variance. We think that the appearance of large trading volume is in large probability due to the new information released to the market. As a result, the volatility of the stock becomes larger in this situation. When we add realized kurtosis to the regression model, it is found from Table $\ref{t11}$ that in the presence of trading volume, realized kurtosis exhibits an extremely significant effect on the future realized variance, while trading volume now becomes insignificant. The reason may be that realized kurtosis contains already some of the information contained in trading volume. One possible explanation is that realized kurtosis measures the jumps within the day which may correspond to the large trading volume for one trade. Consequently, these two measures may have a relationship with each other, which have a correlation of $0.37$. Moreover, the addition of the realized kurtosis improves the $R^2$ from $0.026$ to $0.173$, showing the importance of realized kurtosis.\\
	
	\noindent
	We next consider the effect of positive and negative daily returns towards realized variance. We employ the following equations:
	\begin{equation}
	rvar_{t+1} = \alpha_0 + \alpha_1 dret_t^+ + \alpha_2 dret_t^- + \epsilon_{t+1}
	\end{equation}
	\begin{equation}
	rvar_{t+1} = \alpha_0 + \alpha_1 dret_t^+ + \alpha_2 dret_t^- + \alpha_3 rkurt_t + \epsilon_{t+1}
	\end{equation}
	\noindent
	The results are shown in Tables $\ref{t12}$ and $\ref{t13}$.\\
	\begin{table}[H]
		\centering
		\caption{Regression model (25) for realized variance with previous day signed daily returns}	
		\begin{tabular}{l*{4}{c}}
			\hline
			& $\alpha_1(dret^+)$ & $\alpha_2(dret^-)$ & $R^2$ & F test\\
			\hline
			Estimate  & 3.43$\times 10^{-4}$  & -8.39$\times 10^{-4}$ & 0.033 &    \\
			Standard Error  & 2.73$\times 10^{-4}$ & 3.23$\times 10^{-4}$ &  &   \\
			p-value & 0.21 & 0.01$*$ &  &  0.03  \\
			\hline		
		\end{tabular}
		\label{t12}
	\end{table}
	\begin{table}[H]
		\centering
		\caption{Regression model (26) for realized variance with previous day signed daily returns and realized kurtosis}
		\begin{tabular}{l*{5}{c}}
			\hline
			& $\alpha_1(dret^+)$ & $\alpha_2(dret^-)$ & $\alpha_3(rkurt)$  & $R^2$ & F test\\
			\hline
			Estimate  & -3.79$\times 10^{-4}$ & -4.01$\times 10^{-4}$ & 2.90$\times 10^{4}$  & 0.199 &    \\
			Standard Error  & 2.74$\times 10^{-4}$ & 3.03$\times 10^{-4}$ & 4.56$\times 10^{3}$  &  &   \\
			p-value & 0.17 & 0.19 & 1.41$\times 10^{-9}***$ & & 1.90$\times 10^{-9}$  \\
			\hline		
		\end{tabular}
		
		\label{t13}
	\end{table}
	
	\noindent
	It is found from Table $\ref{t12}$ that positive daily return has no effect while negative daily return has a significant effect on realized variance and the coefficient is negative which indicates clearly the leverage effect mentioned in the literature. When we add realized kurtosis in the regression equation, we can see from Table $\ref{t13}$ that realized kurtosis is extremely significant for realized variance, while the other two measures are now insignificant in explaining realized variance. This indicates realized kurtosis may also contain some information related to the leverage effect. Again, realized kurtosis substantially improves the fitting of the regression, indicated by the increasing $R^2$.\\
	
	\noindent
	Finally, we combine all the pertinent covariates together in regression model (27).
	\begin{equation}
	rvar_{t+1} = \alpha_0 + \alpha_1 rkurt_t + \alpha_2 tvol_t + \alpha_3 dret_t^+ + \alpha_4 dret_t^- + \epsilon_{t+1}
	\end{equation}
	We can see from Table $\ref{t14}$ that realized kurtosis still exhibits an extremely significant effect on the future realized variance.
	\begin{table}[H]
		\centering
		\caption{Regression model (27) for realized variance with all covariates}
		\begin{tabular}{l*{6}{c}}
			\hline
			& $\alpha_1(rkurt)$ & $\alpha_2(tvol)$ & $\alpha_3(dret^+)$ & $\alpha_4(dret^-)$ &  $R^2$ & F test\\
			\hline
			Estimate  & 2.40$\times 10^{4}$ & 1.28$\times 10^{-12}$ & -2.59$\times 10^{-4}$ & -4.00$\times 10^{-4}$ & 0.10 &    \\
			Standard Error  & 6.92$\times 10^{3}$ & 1.17$\times 10^{-12}$ & 3.20$\times 10^{-4}$ & 3.43$\times 10^{-4}$ &  &   \\
			p-value & 6.47$\times 10^{-4}***$ & 0.28 & 0.42 & 0.24 & & 3.20$\times 10^{-4}$  \\
			\hline		
		\end{tabular}
		\label{t14}
	\end{table}

	\subsubsection{Out-of-sample forecasting}
	\noindent
	We investigate whether adding realized kurtosis into the regressive models improves the out-of-sample forecasting accuracy. We focus on the following three regression models: 
	\begin{equation}
	rvar_{t+1} = \alpha_0 + \alpha_1 dret_t + \alpha_2 rskew_t  + \epsilon_{t+1}
	\end{equation}
	\begin{equation}
	rvar_{t+1} = \alpha_0 + \alpha_1 tvol_t + \epsilon_{t+1}
	\end{equation}
	\begin{equation}
	rvar_{t+1} = \alpha_0 + \alpha_1 dret_t^+ + \alpha_2 dret_t^-  + \epsilon_{t+1}
	\end{equation}
	
	\noindent
	After adding the realized kurtosis, the regression models become:
	\begin{equation}
	rvar_{t+1} = \alpha_0 + \alpha_1 dret_t + \alpha_2 rskew_t + \alpha_3 rkurt_t + \epsilon_{t+1}
	\end{equation}
	\begin{equation}
	rvar_{t+1} = \alpha_0 + \alpha_1 tvol_t + \alpha_2 rkurt_t + \epsilon_{t+1}
	\end{equation}
	\begin{equation}
	rvar_{t+1} = \alpha_0 + \alpha_1 dret_t^+ + \alpha_2 dret_t^-  + \alpha_3 rkurt_t + \epsilon_{t+1}
	\end{equation}
	
	\noindent
	We compare the out-of-sample prediction performance of Model (28) against Model (31), Model (29) against Model (32), and Model (30) against Model (33). We use two metrics to do the comparison, the normalized mean square error (MSE) and the Clark and McCraken (CM) test. The normalized MSE is defined as: \\
	\begin{displaymath}
	MSE = \frac{\sum (\text{predicted realized variance} - \text{true value})^2}{\sum (\text{true value})^2}
	\end{displaymath}
	Moreover, the CM statistic refers to the \citet{CM01} Encompassing test, which compares the out-of-sample prediction ability of nested models. The larger the CM statistic, the better the latter model is. The result is shown in Table $\ref{t15}$. The column with $MSE_1$ exhibits the MSE's of Model (31), (32) and (33), while the column with $MSE_2$ for Model (28), (29) and (30).  The last three columns give the 90th, 95th and 99th percentile of the distribution of the statistic derived under the null, which is from Clark and McCraken (2001) and can be treated as the critical values. The sample period is the first 200 days, from January 3, 2012 to October 16, 2012, and the forecast period is the next 40 days, from October 17, 2012 to December 14, 2012.\\
	\begin{table}[H]
		\centering
		\caption{The comparison of the out-of-sample prediction performance}	
		\begin{tabular}{l*{6}{c}}
			\hline
			& $MSE_1$ & $MSE_2$ & CM statistic & 0.90 & 0.95 & 0.99  \\
			\hline
			& \multicolumn{6}{c}{Model (28) and Model (31)} \\
			\hline
			(31) versus (28)  & 0.15 & 0.16 & 2.42 & 0.449 & 0.698 & 1.300   \\
			\hline
			& \multicolumn{6}{c}{Model (29) and Model (32)} \\
			\hline
			(32) versus (29)  & 0.15 & 0.16 & 0.37 & 0.449 & 0.698 & 1.300   \\
			\hline
			& \multicolumn{6}{c}{Model (30) and Model (33)} \\
			\hline
			(33) versus (30)  & 0.16 & 0.16 & 2.52 & 0.449 & 0.698 & 1.300   \\
			\hline		
		\end{tabular}
		\label{t15}
	\end{table}

	\noindent
	From Table $\ref{t15}$, we see that the models including realized kurtosis have less or equal MSE than the ones without realized kurtosis. For instance, in the comparison of Model (29) and Model (32), Model (32) with realized kurtosis has an MSE of $0.15$, which is smaller than $0.16$ of Model (29). Consequently, realized kurtosis helps producing lower forecasting errors. Furthermore, the out-of-sample performance of Models (31) and (33) is significantly better than Models (28) and (30), respectively. In the statistical sense, Model (32) is not significantly better than Model (29), possibly because some information in the realized kurtosis has been taken care of by the trading volume. Additionally, all the prediction errors can be treated as white noise, as the p-values of the Ljung-Box test applied to the prediction errors of the six models above are $0.52$, $0.55$, $0.53$, $0.72$, $0.59$, $0.92$, respectively.\\
	
	\subsection{What about including the past history of realized variance?}
	\noindent
	We employ the following regression models to see whether realized kurtosis still maintains some explaining power when the first lag of realized variance is included:
	
	\begin{equation}
	rvar_{t+1} = \alpha_0 + \alpha_1 rvar_t  + \epsilon_{t+1}
	\end{equation}
	\begin{equation}
	rvar_{t+1} = \alpha_0 + \alpha_1 rvar_t + \alpha_2 rkurt_t + \epsilon_{t+1}
	\end{equation}
	
	\begin{table}[H]
		\centering
		\caption{The comparison of the in-sample and out-of-sample prediction performance}
		\begin{tabular}{l*{6}{c}}
			\hline
			& \multicolumn{6}{c}{In-sample analysis: Regression model (34)} \\
			\hline
			& $\alpha_1(rvar)$ &  & $R^2$ & F test & & \\
			\hline
			Estimate  & 3.83$\times 10^{-1}$ &  & 0.144 & & &  \\
			Standard Error  & 6.64$\times 10^{-2}$  &  &  & & & \\
			p-value & 3.08$\times 10^{-8}***$  &  & &  3.08$\times 10^{-8}$ & & \\
			\hline
			& \multicolumn{6}{c}{In-sample analysis: Regression model (35)} \\
			\hline
			& $\alpha_1(rvar)$ & $\alpha_2(rkurt)$ & $R^2$ & F test & & \\
			\hline
			Estimate  & 7.43$\times 10^{-2}$ & 2.28$\times 10^{4}$ & 0.174 & & &  \\
			Standard Error  & 1.32$\times 10^{-1}$  & 8.48$\times 10^{3}$ &  & & & \\
			p-value & 0.58  & 7.88$\times 10^{-3}**$ & &  6.66$\times 10^{-9}$ & & \\
			\hline
			& \multicolumn{6}{c}{Out-of-sample prediction} \\
			\hline
			& $MSE_1$ & $MSE_2$ & CM statistic & 0.90 & 0.95 & 0.99  \\
			\hline
			(35) versus (34)  & 0.16 & 0.16 & 0.49 & 0.449 & 0.698 & 1.300   \\
			\hline
		\end{tabular}
		
		\label{t16}
	\end{table}
	
	\noindent
	We see from Table $\ref{t16}$ that in Equation (35), realized kurtosis shows a significant effect while the first lag of realized variance exhibits no significant effect. The addition of realized kurtosis improves the $R^2$ by about $3\%$. Furthermore, the emcompassing test shows that the regression model including realized kurtosis marginally outperforms the one without kurtosis. In addition, we include five lags of realized variance in Equations (34) and (35), and find the conclusion is very similar that realized kurtosis has explanatory power for the one-day ahead realized variance in the presence of the five lags of realized variance. Details are omitted here. Therefore, from both in-sample and out-of-sample analysis, realized kurtosis indicates some additional information besides the past history of realized variance.\\
	
	\subsection{Conclusions}
	\noindent
	From the above in-sample and out-of-sample analysis, we conclude that realized kurtosis does have some forecasting power for the future daily volatility within a short horizon. In addition, the in-sample improvement is larger than the out-of-sample one with realized kurtosis, which indicates that the linear regression may not be well suited for nonlinear relationships. Moreover, we see that the $R^2$ for the regression model is small, which indicates that in practice, it is difficult to do the prediction very precisely. That's why the addition of realized kurtosis seems not generating great improvement in the out-of-sample analysis. In the literature, we also witness low $R^2$ in the regression analysis for this field, for example, \citet{BZ06} and \citet{LWL13}. However, the significance of the variable, realized kurtosis, is still valid in this case. \\
	
	\noindent
	Furthermore, we use the stepwise regression with the Akaike information criterion (AIC) to choose the best regressors for the one-day ahead realized variance. We use positive daily return, negative daily return, realized skewness, realized kurtosis and trading volume as regressors for initiation. The resulting regressors are positive daily return and realized kurtosis, which also indicates the significance of realized kurtosis. We conduct the same procedure to those $50$ stocks in Section 4.\\
	
	\section{Robustness of the result}
	\noindent
	From the above empirical analysis, we find that realized kurtosis possesses explaining power for the future daily volatility of the stock IBM, which is by proxy of realized variance. In this section, we explore the performance of realized kurtosis for more stocks and give some explanations for the explaining power.\\
	
	\noindent
	To check whether the performance of realized kurtosis is robust across different stocks, we employ 50 stocks from NYSE, where 10 of small capitalization size, 10 of medium size and 30 of large size. The stocks are the following: 
	\begin{enumerate}
		\item Small: QIHU, FMC, TE, EGO, NNN, BEE, TRN, DKS, SDRL and RAD.
		\item Medium: DV, DSX, TDW, MPW, CIM, MDW, GTI, DF, BTX and DGI.
		\item Large: AIG, AXP, BA, C, CAT, CVX, DD, DIS, GE, GS, HD, HON, IMB, JNJ, JPM, KO, MCD, MMM, MRK, NKE, PFE, PG, T, TRV, UNH, UTX, V, VZ, WMT and XOM.
	\end{enumerate}
	The details of these stock variables can be found in Table \ref{stockinformation} in the appendix. These stocks are from different sectors of the industry, namely technology, basic materials, utilities, financial, services and so on, thus providing sufficient samples of the stocks in the USA market. Note that the tables in this Section are all very long so that we put them in the appendix, making it easy to read.\\
	
	\noindent
	To explore the forecasting power of realized kurtosis on the daily volatility, we adopt the regression models in Section 3, namely Equations (21), (24), (26) and (27). In addition, we use the stepwise method with the AIC to select adequate covariates for the regression of the future realized variance.\\
	
	\noindent
	In Table \ref{resultrk}, we show the result for the performance of realized kurtosis. The first column exhibits the stock variables. The second to fifth columns show the significance of the coefficient for realized kurtosis in Equation (21), (24), (26) and (27), respectively. For example, the second entry in the first row, $dret+rskew$, indicates that the second column shows the forecasting performance of realized kurtosis in the presence of daily return and realized skewness. The numbers shown in the table stand for the different level of significance. "$0$" stands for not significant, with p-value bigger than $0.1$; "$0.5$" stands for marginally significant, with p-value between $0.05$ and $0.1$; "1" stands for significant, with p-value between $0.01$ and $0.05$; "2" stands for very significant, with p-value between $0.001$ and $0.01$; and "3" stands for extremely significant, with p-value less than $0.001$. The last column indicates the result of the covariate selection. The potential covariates include realized skewness, realized kurtosis, trading volume and positive and negative daily returns. The number "$1$" in some entries of the last column means that no covariate is selected.\\
	
	\noindent
	It is found from Table $\ref{resultrk}$ that in general realized kurtosis always shows significant explaining power when trading volume is absent. When trading volume is added, realized kurtosis sometimes exhibits no significant power in forecasting the volatility. We mention above that this may be because realized kurtosis and trading volume share some common information. Furthermore, for the small capitalization group in the covariate selection, 7 of 10 stocks include the realized kurtosis, 5 of 10 stocks include trading volume and only 1 of 10 includes neither of them. For the medium size group, 6 of 10 stocks include realized kurtosis as a covariate, 5 of 10 include trading volume and only 1 excludes both of them. For large size companies, 19 of 30 stocks include realized kurtosis, 14 of 30 include trading volume and 3 of 30 include neither of this two covariates. The performance of realized kurtosis and trading volume is stable, with respect to stocks with different sizes and different sectors.\\
	
	\noindent
	In the above regression analysis, we use realized kurtosis as a covariate for realized variance. However, the orders of the two variables are not the same from Equation (18). As a result, we take the square root of realized kurtosis to make it of the same order with realized variance, and then conduct the same regression analysis. The result is shown in Table $\ref{resultsqrtrk}$. Remember that all $rkurt$'s in Table $\ref{resultsqrtrk}$ stand for the square root of realized kurtosis.\\
	
	\noindent
	It is found that the square root of realized kurtosis performs better in explaining the future daily volatility, as shown by Columns 2 to 5 of Table \ref{resultsqrtrk}. When all possible covariates are counted in, the square root of realized kurtosis shows a significant effect on the future realized variance in 38 of the 50 cases. Furthermore, for the small capitalization group in the covariate selection, 7 of 10 stocks include realized kurtosis, 4 of 10 stocks include trading volume and only 1 of 10 includes neither of them. For the medium size group, 9 of 10 stocks include realized kurtosis as a covariate, 4 of 10 include trading volume and no stocks excludes neither of them. For large size companies, 25 of 30 stocks include realized kurtosis, 10 of 30 include trading volume and only 1 of 30 includes neither of this two covariates. We see that the performance of the square root of realized kurtosis is even better than that of realized kurtosis, especially for the firms with medium and large sizes. In addition, it seems that the square root of realized kurtosis and trading volume are able to account for almost all explaining power for the future daily volatility, in general.\\
	
	\noindent
	From Equation (18), realized variance converges to the sum of two parts, the integrated variance and the sum of the square of jumps. Additionally, realized kurtosis converges to the sum of the fourth power of price jumps. As a consequence, it is natural to wonder if realized kurtosis possesses some explaining power for realized variance as they both have a jump component. However, this seems not to be the case. In \citet{HT05}, the authors separate the two components of realized variance to check for the contribution of the jump component. In the empirical study, they find that the jump component only accounts for $7\%$ of stock market price variance, which indicates that it is the continuous component that dominates. As a result, why the previous-day price jump affects the continuous price fluctuation is worth considering. In addition, even if the jump part is really important for daily volatility, we know that the jump of the price always corresponds to the unexpected arrival of new information, so that it is unnatural that the previous-day jump has strong forecasting power on future's jumps. Nevertheless, we separate out the continuous component of realized variance, estimated by the bipower variation. We conduct all the regression models with respect to the bipower variation to see whether the explanatory power of the square root of realized kurtosis remains. The result is shown in Table $\ref{resultsqrtrkbv}$.\\

	\noindent
	It is found that on average, the square root of realized kurtosis remains powerful in predicting the future daily bipower variation, shown by Columns 2 to 5 of Table \ref{resultsqrtrkbv}. However, the performance is slightly worse than that under realized variance, which is also shown in the covariate selection column. For stocks with small capitalization, 7 of 10 stocks include realized kurtosis, 6 of 10 stocks include trading volume and only 2 of 10 include neither of them. For the medium size, 7 of 10 stocks include realized kurtosis as a covariate, 5 of 10 include trading volume and only 1 excludes both of them. For large size companies, 23 of 30 stocks include realized kurtosis, 11 of 30 include trading volume and 3 of 30 include neither of this two covariates.\\
	
\begin{figure}[H]
		\centering
		\includegraphics[width=11cm]{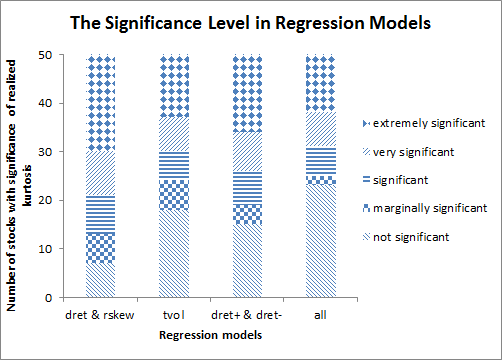}
		\caption{The forecasting power of realized kurtosis towards realized variance in regression models}
		\label{f11}
	\end{figure}

\begin{figure}[H]
		\centering
		\includegraphics[width=11cm]{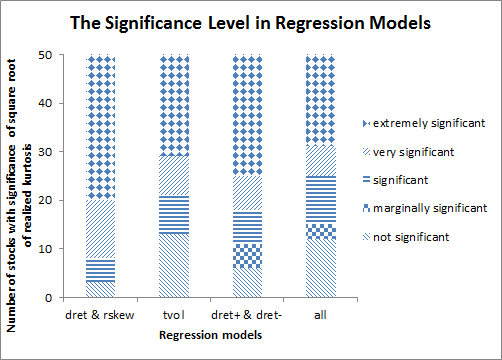}
		\caption{The forecasting power of the square root of realized kurtosis towards realized variance in regression models}
		\label{f12}
	\end{figure}

\begin{figure}[H]
		\centering
		\includegraphics[width=11cm]{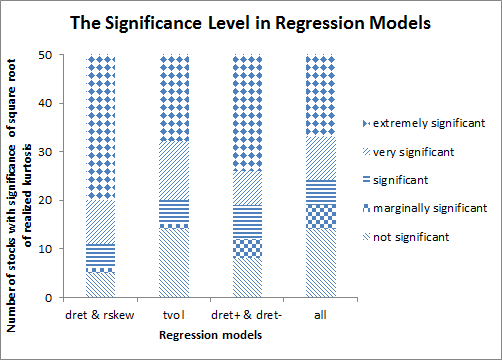}
		\caption{The forecasting power of the square root of realized kurtosis towards bipower variation in regression models}
		\label{f13}
	\end{figure}

\begin{figure}[H]
		\centering
		\includegraphics[width=12cm]{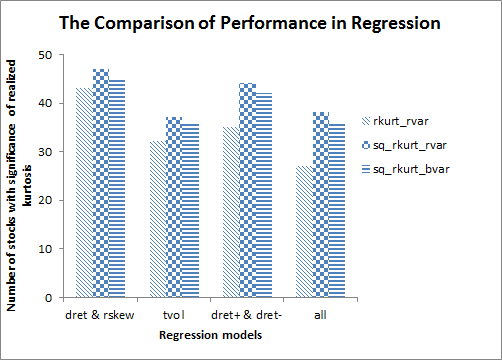}
		\caption{The comparison of the forecasting power for different regression models}
		\label{f14}
	\end{figure}

	\noindent
	In summary, we use the following graphs to illustrate the overall performance of the regressors. In Figures \ref{f11}-\ref{f13}, each bar represents the forecasting performance in one regression model involving the variables shown below the bar. Different filling patterns stand for the level of significance for realized kurtosis or the square root of realized kurtosis, shown by the legend, for example, horizontal line means realized kurtosis is significant, with p-value between $0.01$ and $0.05$. Figure \ref{f14} compares the performance of realized kurtosis forecasting realized variance, the square root of realized kurtosis forecasting realized variance, and the square root of realized kurtosis forecasting the bipower variation. 
	
	\noindent
	From the graphs, we see that the (square root of) realized kurtosis always performs better in the absence of trading volume. Nevertheless, the overall performance of the (square root of) realized kurtosis is satisfactory. The worst case is when we add all variables in the regression models for realized kurtosis, where about $60\%$ of the stocks still indicate the significance of realized kurtosis forecasting the future realized variance. From Figure \ref{f14}, we see that the performance of the square root of realized kurtosis forecasting realized variance is always the best among the three no matter what the regression model is.\\
	
	\noindent
	It appears that realized kurtosis, whether taking the square root or not, measures the price jump size to some extent. It contains information for the future daily volatility, possibly due to the following reasons. Firstly, \citet{M76} points out that the continuous part of the stock volatility may be due to the change in the economic anticipation and the temporary imbalance between supply and demand. Moreover, sometimes the price jumps are incidental, which correspond to the newly arrived news to the market. When the market is unable to digest the news efficiently, the news effect aggregates and the jump should have some forecasting power for the future volatility. This is also indicated by the result that the long horizon forecasting performance of realized kurtosis becomes worse since the news has been digested gradually by the market after a long time. Additionally, sometimes the price jump is artificial, i.e. the price jump is due to manipulation by some large financial institutions. In this case, the market will fluctuate corresponding to the reaction of the public and partly the follow-up actions by the institutions. Consequently, the future volatility has some relationships with previous price jumps. Certainly, there remain some other reasons, undiscovered.\\
	
	\section{Conclusions}
	\noindent
	In this paper, we analyse whether higher realized moments have explaining power for future daily returns or realized variance. We find that realized skewness provides not enough evidence of the forecasting power for daily returns, in contrast with the literature. On the other hand, realized kurtosis exhibits significant forecasting power for the future realized variance in short period. Furthermore, the square root of realized kurtosis shows even better forecasting ability.\\
	
	\noindent
	However, it is found that in the regression analysis, the $R^2$ is low, even though the effect of realized kurtosis is significant, which indicates that the prediction cannot be very precise when linear regression models are used. This phenomenon also suggests that some nonlinear regression models may be used to fit the relationships between realized variance and realized kurtosis, which will be pursued in the future.\\
	\\

	\newpage
	\begin{appendices}
		\setcounter{table}{0}
		\renewcommand{\thetable}{A.\arabic{table}}
		
	\begin{center}
		\begin{small}
			\begin{longtable}[H]{|l|c|c|c|}
				\caption{Information of the selected stocks}\\
				\caption*{{\footnotesize In the Sector column, Tech stands for Technology, Bmat for Basic materials, Util for Utilities, Fin for Financial, Serv for Services, Igood for Industrial goods, Cgood for Consumer goods, and Heal for Healthcare.}}\\
				
				\hline
				Stock & Firm & Sector & Industry \\
				\hline
				QIHU & Qihoo 360 Technology & Tech & Information technology/services\\
				FMC	& FMC &	Bmat &	Chemicals\\
				TE & TECO Energy	& Util	& Electric\\
				EGO	& Eldorado Gold &	Bmat	& Gold\\
				NNN	& National Retail Properties &	Fin &	REIT\\
				BEE	& Strategic Hotels $\&$ Resorts &	Fin &	REIT\\
				TRN	& Trinity Industries	& Serv	& Railroads\\
				DKS	& Dick's Sporting Goods	& Serv	& Sporting goods stores\\
				SDRL	& SeaDrill	& Bmat	& Oil $\&$ gas drilling $\&$ exploration\\
				RAD	& Rite Aid	& Serv	& Drug stores\\
				\hline
				DV	& DeVry Education	& Serv	& Education $\&$ training services\\
				DSX	& Diana Shipping & Serv	& Shipping\\
				TDW	& Tidewater	& Bmat	& Oil $\&$ gas equipment $\&$ services\\
				MPW	& Medical Properties Trust	& Fin	& REIT\\
				CIM	& Chimera Investment	& Fin	& REIT\\
				MDW	& Midway Gold	& Bmat	& Gold\\
				GTI	& GrafTech International	& Igood	& Industrial electrical equipment\\
				DF	& Dean Foods	& Cgood	& Dairy products\\
				BTX	& Bio Time	& Heal	& Biotechnology\\
				DGI	& DigitalGlobe	& Igood	& Aerospace/defense products $\&$ services\\
				\hline
				AIG	& American International Group	& Fin	& Property $\&$ casualty insurance\\
				AXP	& American Express	& Fin	& Credit services\\
				BA	& Boeing	& Igood	& Aerospace/defense products $\&$ services\\
				C	& Citigroup	& Fin	& Money center banks\\
				CAT	& Catepillar & Igood	& Farm $\&$ construction machinery\\
				CVX	& Chevron	& Bmat	& Major integrated oil $\&$ gas\\
				DD	& E.I. du Pont de Nemours	& Bmat	& Agricultural chemicals\\
				DIS	& Walt Disney	& Serv	& Entertainment\\
				GE	& General Electric	& Igood	& Diversified machinery\\
				GS	& Goldman Sachs	& Fin	& Investment brokerage\\
				HD	& Home Depot	& Serv	& Home improvement stores\\
				HON	& Honeywell International	& Igood	& Diversified machinery\\
				IBM	& International Business Machines	& Tech	& Information technology services\\
				JNJ	& Johnson $\&$ Johnson	& Heal	& Drug manufacturers\\
				JPM	& JPMorgan	& Fin	& Money center banks\\
				KO	& Coca-Cola	& Cgood	& Beverages-soft drinks\\
				MCD	& McDonald's	& Serv	& Restaurants\\
				MMM	& 3M	& Igood	& Diversified machinery\\
				MRK	& Merch $\&$ Co	& Heal	& Drug manufacturers\\
				NKE	& NIKE	& Cgood	& Textile-apparel footwear $\&$ accessories\\
				PFE	& Pfizer &	Heal	& Drug manufacturers\\
				PG	& Procter $\&$ Gamble	& Cgood	& Personal products\\
				T	& AT$\&$T	& Tech	& Telecom services\\
				TRV	& Travelers & Fin	& Property $\&$ casualty insurance\\
				UNH	& UnitedHealth	& Heal	& Health care plans\\
				UTX	& United Technologies	& Igood	& Aerospace/defense products $\&$ services\\
				V	& Visa	& Fin	& Credit services\\
				VZ	& Verizon Communications	& Tech	& Telecom services\\
				WMT	& Wal-Mart Stores	& Serv	& Discount, variety stores\\
				XON	& Intrexon	& Heal	& Biotechnology\\
				\hline				
				\label{stockinformation}
			\end{longtable}
		\end{small}
	\end{center}

	\newpage
	\begin{center}
		\begin{longtable}[H]{|l|c|c|c|c|c|}			
			\caption{The performance of realized kurtosis towards realized variance}\\
			\caption*{{\footnotesize The second to fifth columns show the significance of the coefficient for realized kurtosis in Equation (21), (24), (26) and (27), respectively. The numbers shown in the table stand for the different level of significance. "$0$" stands for not significant, with p-value bigger than $0.1$; "$0.5$" stands for marginally significant, with p-value between $0.05$ and $0.1$; "1" stands for significant, with p-value between $0.01$ and $0.05$; "2" stands for very significant, with p-value between $0.001$ and $0.01$; and "3" stands for extremely significant, with p-value less than $0.001$. The last column indicates the result of the covariate selection. The potential covariates include realized skewness, realized kurtosis, trading volume and positive and negative daily returns. The number "$1$" in some entries of the last column means that no covariate is selected.}}\\
			\hline
			Stock & $dret$ & $tvol$ & $dret^+$ & all & covariate selection \\
			& $\& rskew$ & & $\& dret^-$ & & \\
			\hline
			QIHU	& 0	& 0	& 0	& 0	& $tvol$\\
			FMC	& 3	& 2	& 3	& 1	& $tvol+rkurt$\\
			TE	& 1	& 2	& 2	& 2	& $dret^-+rkurt$\\
			EGO	& 3	& 0	& 0.5	& 0	& $dret^-+rskew+rkurt+tvol$\\
			NNN& 0.5& 2	& 2	& 2	& $rkurt$\\
			BEE	& 3	& 2	& 3	& 3	& $rkurt$\\
			TRN	& 3	& 3	& 3	& 3	& $rskew+rkurt$\\
			DKS	& 1	& 0	& 0	& 0	& $dret^-+tvol$\\
			SDRL& 2	& 1	& 2	& 1	& $rkurt+tvol$\\
			RAD	& 1	& 0	& 0	& 0	& $dret^++dret^-$\\
			\hline
			DV	& 2	& 0	& 1	& 0	& $tvol$\\
			DSX	& 3	& 3	& 3	& 3	& $dret^++dret^-+rkurt$\\
			TDW	& 0	& 0	& 0	& 0	& $dret^++tvol$\\
			MPW	& 1	& 3	& 2	& 2 	& $rkurt$\\
			CIM	& 2	& 0.5	& 1	& 0	& $dret^++dret^-+rkurt+tvol$\\
			MDW	& 3	& 3	& 3	& 3	& $dret^-+rkurt+tvol$\\
			GTI	& 0	& 0	& 0	& 0	& $dret^-$\\
			DF & 0.5& 0	& 0	& 0	& $dret^++tvol$\\
			BTX	& 3	& 1	& 0.5	& 0	& $dret^-+rkurt$\\
			DGI	& 1	& 0	& 0	& 0	& $rskew+rkurt$\\
			\hline
			AIG	& 3	& 3	& 3	& 3	& $dret^-+rskew+rkurt$\\
			AXP	& 0	& 0	& 0	& 0	& $1$\\
			BA& 0.5	& 0.5	& 0	& 0	& $dret^++dret^-$\\
			C	& 2	& 0	& 1	& 0	& $tvol$\\
			CAT	& 3	& 3	& 3	& 3	& $dret^-+rkurt$\\
			CVX	& 3	& 3	& 3	& 3	& $rkurt+tvol$\\
			DD& 0.5	& 2	& 2	& 2	& $dret^-+rkurt+tvol$\\
			DIS	& 3	& 0	& 2	& 0	& $dret^++tvol$\\
			GE	& 3	& 3	& 3	& 2	& $dret^-+rkurt+tvol$\\
			GS	& 3	& 1	& 3	& 1	& $rskew+rkurt+tvol$\\
			HD	& 2	& 0.5	& 2	& 0	& $dret^++tvol$\\
			HON	& 3	& 3	& 3	& 3	& $rkurt$\\
			IBM	& 3	& 3	& 3	& 3	& $dret^++rkurt$\\
			JNJ	& 0	& 0	& 0.5	& 0	& $tvol$\\
			JPM	& 1	& 0	& 0	& 0	& $tvol$\\
			KO	& 3	& 3	& 3	& 3	& $dret^++rkurt$\\
			MCD	& 1	& 0	& 0	& 0	& $dret^-+rskew+rkurt$\\
			MMM	& 2	& 2	& 2	& 2	& $dret^-+rkurt$\\
			MRK	& 1	& 1	& 0.5	& 1	& $rskew+rkurt$\\
			NKE	& 3	& 0	& 1	& 0	& $rskew+rkurt$\\
			PFE& 0.5& 0.5	& 1	& 0.5	& $dret^-+rskew+rkurt+tvol$\\
			PG	& 3	& 2	& 1	& 1	& $rkurt$\\
			T	& 3	& 3	& 3	& 3	& $rkurt$\\
			TRV& 0.5& 0.5	& 0	& 0	& $dret^-+tvol$\\
			UNH	& 0	& 0	& 0	& 0	& $1$\\
			UTX	& 0	& 1	& 1	& 1	& $rkurt$\\
			V	& 2	& 0	& 0	& 0	& $dret^-+rskew+tvol$\\
			VZ	& 3	& 3	& 3	& 3	& $rskew+rkurt$\\
			WMT	& 2	& 0.5	& 0	& 0	& $dret^++dret^-+tvol$\\
			XON	& 2	& 1	& 3	& 0.5	& $dret^-+rkurt+tvol$\\
			\hline				
			\label{resultrk}
		\end{longtable}
		
	\end{center}
	
	\newpage
	\begin{center}
		\begin{longtable}[H]{|l|c|c|c|c|c|}
			\caption{The performance of the square root of realized kurtosis towards realized variance}
			\\
			\hline
			Stock & $dret$ & $tvol$ & $dret^+$ & all & covariate selection \\
			& $\& rskew$ & & $\& dret^-$ & & \\
			\hline
			QIHU& 0	& 0	& 0	& 0	& $tvol$\\
			FMC	& 3	& 3	& 3	& 3	& $tvol+rkurt$\\
			TE	& 1	& 2	& 2	& 2	& $dret^-+rkurt$\\
			EGO	& 3	& 1	& 2	& 1	& $dret^-+rs+rkurt$\\
			NNN	& 2	& 3	& 3	& 3	& $rkurt$\\
			BEE	& 3	& 3	& 3	& 3	& $rkurt$\\
			TRN	& 3	& 3	& 3	& 3	& $rskew+rkurt$\\
			DKS	& 3	& 0	& 0.5	& 0	& $dret^-+tvol$\\
			SDRL& 3	& 3	& 3	& 3	& $rkurt+tvol$\\
			RAD	& 1	& 0	& 0	& 0	& $dret^++dret^-$\\
			\hline
			DV	& 3	& 0	& 3	& 2	& $dret^++dret^-+rskew+rkurt+tvol$\\
			DSX	& 3	& 3	& 3	& 3	& $dret^++dret^-+rkurt$\\
			TDW	& 0	& 0& 0.5& 0	& $dret^++rskew+rkurt+tvol$\\
			MPW	& 3	& 3	& 3	& 3	& $rkurt$\\
			CIM	& 3	& 2	& 2	& 1	& $dret^++dret^-+rkurt$\\
			MDW	& 3	& 3	& 3	& 3	& $dret^-+rkurt+tvol$\\
			GTI	& 2	& 2& 0.5& 1	& $rkurt$\\
			DF	& 2	& 0	& 0	& 0	& $dret^++tvol$\\
			BTX	& 3	& 3	& 2	& 2	& $dret^-+rkurt$\\
			DGI	& 2	& 0	& 1	& 0.5	& $rskew+rkurt$\\
			\hline
			AIG	& 2	& 3	& 3	& 3	& $dret^-+rskew+rkurt$\\
			AXP	& 1	& 1	& 1	& 1	& $rkurt$\\
			BA	& 1	& 1& 0.5& 0.5	& $dret^++dret^-+rkurt$\\
			C	& 2	& 0	& 2	& 0	& $tvol$\\
			CAT	& 3	& 3	& 3	& 3	& $dret^-+rkurt$\\
			CVX	& 3	& 3	& 3	& 3	& $rkurt$\\
			DD	& 2	& 1	& 1	& 1	& $dret^-+rkurt$\\
			DIS	& 3	& 1	& 3	& 0	& $dret^++tvol+rskew+rkurt$\\
			GE	& 3	& 3	& 3	& 3	& $rkurt$\\
			GS	& 3	& 2	& 3	& 2	& $rskew+rkurt+tvol$\\
			HD	& 3	& 3	& 3	& 2	& $dret^++rkurt+tvol$\\
			HON	& 3	& 3	& 3	& 3	& $rkurt$\\
			IBM	& 3	& 3	& 3	& 3	& $rkurt$\\
			JNJ	& 3	& 1	& 3	& 1	& $rkurt$\\
			JPM	& 2	& 0	& 1	& 0	& $tvol$\\
			KO	& 3	& 3	& 3	& 3	& $dret^++rkurt$\\
			MCD	& 2	& 0	& 0	& 0	& $dret^-+rskew+rkurt$\\
			MMM	& 3	& 2	& 2	& 2	& $rkurt$\\
			MRK	& 2	& 2	& 1	& 1	& $rskew+rkurt$\\
			NKE	& 3	& 3	& 3	& 3	& $rskew+rkurt$\\
			PFE	& 1	& 1	& 2	& 1	& $dret^++rskew+rkurt+tvol$\\
			PG	& 3	& 3	& 3	& 3	& $rkurt$\\
			T	& 3	& 3	& 3	& 3	& $rkurt$\\
			TRV	& 2	& 1	& 1	& 0.5	& $dret^-+rkurt+tvol$\\
			UNH	& 0	& 0	& 0	& 0	& $1$\\
			UTX	& 2	& 2	& 1	& 1	& $rkurt$\\
			V	& 3	& 0	& 0	& 0	& $dret^++tvol$\\
			VZ	& 3	& 3	& 3	& 3	& $rskew+rkurt$\\
			WMT	& 3	& 0& 0.5& 0	& $dret^++dret^-+tvol$\\
			XOM	& 3	& 2	& 3	& 1	& $dret^++rkurt+tvol$\\
			\hline				
			\label{resultsqrtrk}
		\end{longtable}
	\end{center}
	
	\newpage
	\begin{center}
		\begin{longtable}[H]{|l|c|c|c|c|c|}
			\caption{The performance of the square root of realized kurtosis with respect to the bipower variation}
			\\
			\hline
			Stock & $dret$ & $tvol$ & $dret^+$ & all & covariate selection \\
			& $\& rskew$ & & $\& dret^-$ & & \\
			\hline 
			QIHU& 0	& 0	& 0	& 0	& $dret^+$\\
			FMC	& 3	& 2	& 3	& 2	& $tvol+rkurt$\\
			TE	& 1	& 1	& 1	& 0.5 & $dret^-+rskew+rkurt$\\
			EGO	& 3	& 2	& 3	& 2	& $dret^-+rskew+rkurt+tvol$\\
			NNN	& 3	& 2	& 2	& 2	& $rskew+rkurt$\\
			BEE	& 3	& 3	& 3	& 3	& $rskew+rkurt+tvol$\\
			TRN	& 3	& 3	& 3	& 3	& $dret^++rskew+rkurt+tvol$\\
			DKS	& 2	& 0& 0.5& 0	& $dret^-+tvol$\\
			SDRL& 3	& 3	& 3	& 3	& $dret^-+rskew+rkurt+tvol$\\
			RAD& 0.5& 0	& 0	& 0	& $dret^-$\\
			\hline
			DV	& 3	& 0	& 3	& 2	& $dret^++dret^-+rskew+rkurt+tvol$\\
			DSX	& 3	& 3	& 3	& 3	& $rskew+rkurt+tvol$\\
			TDW	& 0	& 0	& 1	& 0	& $dret^++tvol$\\
			MPW	& 3	& 3	& 3	& 3	& $dret^-+rkurt$\\
			CIM	& 3	& 2	& 1	& 1	& $dret^++dret^-+rkurt$\\
			MDW	& 3	& 3	& 3	& 3	& $rkurt$\\
			GTI	& 1	& 1	& 0	& 0	& $dret^++dret^-$\\
			DF	& 3	& 0	& 0	& 0	& $dret^++tvol$\\
			BTX	& 3	& 2	& 2	& 0.5	& $dret^++dret^-+rkurt+tvol$\\
			DGI	& 2	& 0	& 1	& 0	& $rskew+rkurt$\\
			\hline
			AIG	& 2	& 2	& 2	& 2	& $dret^-+rskew+rkurt$\\
			AXP	& 0	& 0	& 0	& 0	& $1$\\
			BA	& 2	& 2	& 1	& 1	& $dret+rkurt$\\
			C	& 2	& 0	& 1	& 0	& $tvol+dret^-$\\
			CAT	& 3	& 3	& 3	& 3	& $dret^-+rskew+rkurt$\\
			CVX	& 3	& 3	& 3	& 3	& $rkurt+tvol$\\
			DD	& 2	& 3	& 3	& 3	& $dret^-+rkurt+tvol$\\
			DIS	& 3	& 1	& 3	& 0.5 & $dret^++tvol+rskew+rkurt$\\
			GE	& 3	& 3	& 3	& 3	& $rkurt$\\
			GS	& 3	& 3	& 3	& 3	& $rskew+rkurt+tvol$\\
			HD	& 3	& 2	& 3	& 1	& $dret^++rskew+rkurt+tvol$\\
			HON	& 3	& 3	& 3	& 3	& $rkurt$\\
			IBM	& 3	& 3	& 3	& 3	& $rkurt$\\
			JNJ	& 3& 0.5& 2	& 0.5	& $rskew+rkurt$\\
			JPM	& 1	& 0	& 0	& 0	& $tvol$\\
			KO	& 3	& 3	& 3	& 3	& $dret^++rkurt$\\
			MCD	& 2	& 0	& 0	& 0	& $dret^-+rskew+rkurt$\\
			MMM	& 3	& 2	& 2	& 2	& $rkurt$\\
			MRK	& 2	& 2	& 2	& 2	& $rkurt$\\
			NKE	& 3	& 3	& 3	& 3	& $rskew+rkurt$\\
			PFE	& 0	& 2	& 3	& 2	& $dret^-+rskew$\\
			PG	& 3	& 3	& 2	& 2	& $rkurt$\\
			T	& 3	& 3	& 3	& 3	& $rkurt$\\
			TRV	& 1	& 1& 0.5& 0.5	& $dret^-+tvol$\\
			UNH	& 0	& 0	& 0	& 0	& $1$\\
			UTX	& 1	& 1	& 1	& 1	& $rkurt$\\
			V	& 2	& 0& 0.5& 0	& $dret^-+rskew+rkurt+tvol$\\
			VZ	& 3	& 3	& 3	& 3	& $rskew+rkurt$\\
			WMT	& 3	& 0& 0.5& 0	& $dret^++rskew+tvol$\\
			XOM	& 3	& 2	& 3	& 1	& $dret^++rkurt+tvol$\\
			\hline				
			\label{resultsqrtrkbv}
		\end{longtable}
	\end{center}
			
	\end{appendices}

\end{document}